\documentclass{agujournal}

\journalname{JGR-Space Physics}
\usepackage{hyperref}

\begin{document}

\title{Cosmic Noise Absorption During Solar Proton Events in WACCM-D and Riometer Observations}

\authors{Erkka Heino\affil{1,2}, Pekka T. Verronen\affil{3}, Antti Kero\affil{4}, Niilo Kalakoski\affil{3}, Noora Partamies\affil{1,5}}

\affiliation{1}{Department of Arctic Geophysics, The University Centre in Svalbard, Longyearbyen, Norway}
\affiliation{2}{Department of Physics and Technology, University of Troms\o , Troms\o, Norway}
\affiliation{3}{Space and Earth Observation Centre, Finnish Meteorological Institute, Helsinki, Finland}
\affiliation{4}{Sodankyl\"a Geophysical Observatory, University of Oulu, Sodankyl\"a, Finland}
\affiliation{5}{Birkeland Centre for Space Science, Bergen, Norway}

\correspondingauthor{Erkka Heino}{erkkah@unis.no}

\begin{keypoints}
\item CNA during 62 SPEs was studied with WACCM-D and riometer observations
\item CNA is modeled well on average and in individual events with a non-linearity correction for high levels of CNA
\item The fixed proton cutoff latitude in WACCM-D at $60^\circ$ leads to overestimation of the extent of the CNA, especially in small to moderate SPEs
\end{keypoints}

\begin{abstract}

Solar proton events (SPEs) cause large-scale ionization in the middle atmosphere leading to ozone loss and changes in the energy budget of the middle atmosphere. The accurate implementation of SPEs and other particle ionization sources in climate models is necessary to understand the role of energetic particle precipitation (EPP) in climate variability.
We use riometer observations from 16 riometer stations and the Whole Atmosphere Community Climate Model with added \textit{D}~region ion chemistry (WACCM-D) to study the spatial and temporal extent of cosmic noise absorption (CNA) during 62 solar proton events from 2000 to 2005. We also present a correction method for the non-linear response of observed CNA during intense absorption events.
We find that WACCM-D can reproduce the observed CNA well with some need for future improvement and testing of the used EPP forcing. The average absolute difference between the model and the observations is found to be less than 0.5 dB poleward of about $66^\circ$ geomagnetic latitude, and increasing with decreasing latitude to about 1 dB equatorward of about $66^\circ$ geomagnetic latitude. The differences are largest during twilight conditions where the modeled changes in CNA are more abrupt compared to observations. An overestimation of about $1^\circ$ to $3^\circ$ geomagnetic latitude in the extent of the CNA is observed due to the fixed proton cutoff latitude in the model. An unexplained underestimation of CNA by the model during sunlit conditions is observed at stations within the polar cap during 18 of the studied events.

\end{abstract}

\section{Introduction}

Solar proton events (SPEs) are large, albeit infrequent, expulsions of energetic particles from the Sun that can last from a few hours to multiple days. A SPE is defined as a period of time where the $\geq 10$ MeV integral proton flux, measured by a geosynchronous satellite, exceeds 10 pfu (particle flux unit, cm$^{-2}$s$^{-1}$sr$^{-1}$). The dominant particle species in SPEs is protons, which are accelerated near the Sun to energies of 10 keV/nucl to multiple GeV/nucl~\citep{kallenrode2003} by solar flares and coronal mass ejection (CME) driven shocks~\citep[e.g.,][]{reames1999}. The acceleration processes get their energy from the magnetic energy stored in the solar corona, but the exact acceleration mechanisms are still being discussed~\citep{vainio2009}. High-energy SPE protons and electrons, as well as energetic electrons from the outer radiation belt, have access to the mesosphere and upper stratosphere in the magnetic polar regions affecting the neutral composition and dynamics of the middle atmosphere~\citep{sinnhuber2012,verronen2013}.

Ionization in the middle atmosphere due to energetic particle precipitation (EPP) causes production of odd hydrogen (HO$_x$) and odd nitrogen (NO$_x$) species that lead to the loss of ozone (O$_3$) through catalytic ozone loss cycles. Odd hydrogen species have a short chemical lifetime and an effect on ozone loss in the mesosphere. Odd nitrogen species are destroyed in the sunlit atmosphere and thus have a long lifetime during the polar winter. Due to its long chemical lifetime in the dark atmosphere, NO$_x$ is subject to transport in the middle atmosphere and has an important effect on stratospheric ozone loss~\citep{randall2005}. \citet{funke2014} showed from MIPAS/Envisat observations that EPP-produced reactive reservoir nitrogen species (NO$_y$) descent regularly down into the stratosphere during polar winter. A NO$_2$ increase of several hundred percent and an O$_3$ decrease of tens of percent between 36 and 60 km altitude, due to the SPEs of October--November 2003, was reported by~\citet{seppala2004} based on GOMOS/Envisat observations. This SPE effect on the NO$_2$ and O$_3$ concentrations was observed to last several months after the SPEs. Ozone is the dominant absorber of UV radiation in the atmosphere and therefore important in the energy budget of the middle atmosphere. Changes in ozone concentrations in the stratosphere have been shown to affect ground-level climate variability especially in the polar regions~\citep{gillett2003}. As EPP affects ozone variability in the middle atmosphere, a similar ground-level coupling effect has been suggested and possible ground-level signatures have been observed and modeled~\citep{seppala2009,baumgaertner2011}. The implementation of EPP ionization in climate models is therefore necessary to understand the role of EPP in climate variability on longer time scales~\citep{andersson2014,matthes2017}.

Increased ionization due to EPP causes absorption of high-frequency radio waves in the polar \textit{D}~region, which has been measured with riometers since the 1950s~\citep{little1958,little1959}. Riometers are passive instruments that measure cosmic radio noise continuously, typically at 30 to 40 MHz frequency. Absorption of radio waves, or cosmic noise absorption (CNA), in the ionosphere is determined with riometers from the difference between the measured radio noise and a quiet day curve (QDC), which is the expected level of radio noise without absorption---i.e., during a "quiet" day.

In this paper, we use riometer observations and the Whole Atmosphere Community Climate Model with added \textit{D}~region ion chemistry (WACCM-D) to study the spatial and temporal extent of CNA caused by SPEs, the effect of geomagnetic cutoff on the CNA, and the ability of the WACCM-D model to reproduce the level and time behavior of observed CNA during SPEs. The non-linear response of riometers to high levels of CNA is also presented and discussed.

\section{Observational Data}
\label{sec:obs}

The used observational data cover 62 SPEs from 2000 to 2005, whose occurrence times were taken from the SPE list at \texttt{ftp://ftp.swpc.noaa.gov/pub/indices/SPE.txt} updated by the National Oceanic and Atmospheric Administration (NOAA). The studied 62 SPEs (see Tables~\ref{tab:eventList1} and \ref{tab:eventList2}) were chosen based on availability of data from the Longyearbyen and Kilpisj\"arvi imaging riometers. The data set consists of 34 S1-class (maximum $\geq 10$ MeV integral proton flux  $\geq10$ pfu), 17 S2-class ($\geq100$ pfu), 6 S3-class ($\geq1,000$ pfu), and 5 S4-class ($\geq10,000$ pfu) SPEs. The number of available stations varies within the used events.

\begin{figure}[h]
 \centering
\centerline{\includegraphics[width=9.5cm]{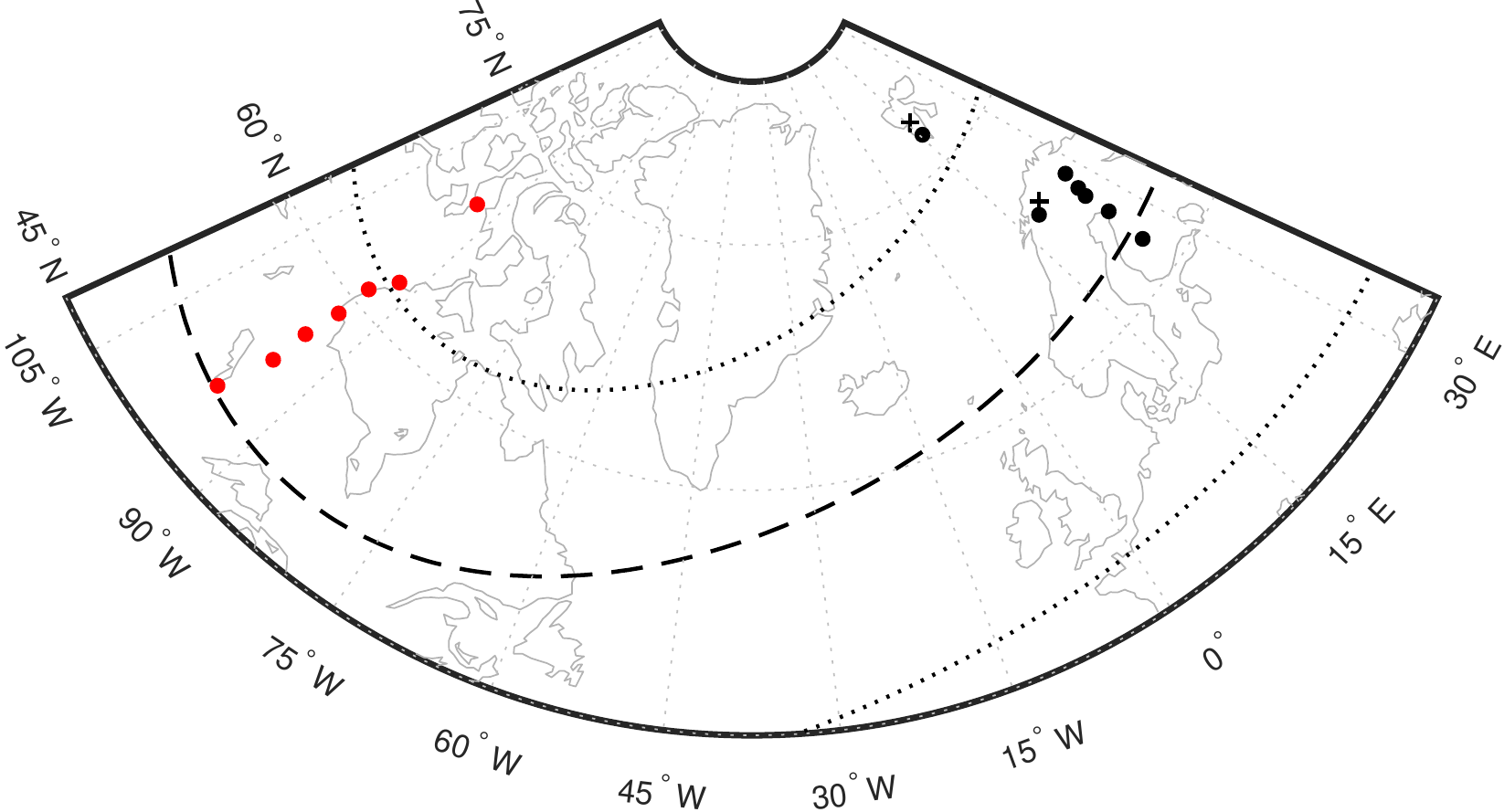}}
 \caption{Geographical locations of the riometers used in the study and the limits of the SPE and medium-energy electron ionization in the used model. The Canadian GO-Canada chain riometers are marked with red dots, the SGO riometers with black dots, and the two imaging riometers with black crosses. The upper and lower limits ($72^\circ$ and $45^\circ$ geomagnetic latitude) of the medium-energy electron ionization are marked with the black dotted lines and the lower limit of the SPE ionization is marked with the black dashed line.}
 \label{fig:stations}
 \end{figure}

The used riometer CNA data are from two arrays of riometers in northern Europe and Canada (See Table~\ref{tab:stations} and Figure~\ref{fig:stations}). The riometers in the European sector are the Sodankyl\"a Geophysical Observatory (SGO) wide-beam riometer network spanning from Jyv\"askyl\"a, Finland to Hornsund, Svalbard, and imaging riometers located at Kilpisj\"arvi, Finland~\citep{browne1995} and Longyearbyen, Svalbard~\citep{stauning1995}. The Canadian sector riometers are the Churchill line stations of the GO-Canada (formerly NORSTAR) wide-beam riometer array~\citep{rostoker1995}.

\begin{table}
\caption{\textit{Names, Locations, and Operating Frequencies of Riometers Used in This Study}}
\centering
\begin{tabular}{lccc}
\hline
Station name and code & Latitude [$^\circ$] & Longitude [$^\circ$] & Frequency [MHz]\\
\hline
European chain: & & &\\
Longyearbyen (LYR) & $75.18^\circ$ ($78.20^\circ$) & $111.11^\circ$ ($15.82^\circ$) & 38.2\\
Hornsund (HOR) & $74.05^\circ$ ($77.00^\circ$) & $108.77^\circ$ ($15.60^\circ$) & 30.0\\
Kilpisj\"arvi (KIL) & $65.82^\circ$ ($69.05^\circ$) & $103.54^\circ$ ($20.79^\circ$) & 38.2\\
Abisko (ABI) & $65.25^\circ$ ($68.40^\circ$) & $101.59^\circ$ ($18.90^\circ$) & 30.0\\
Ivalo (IVA) & $65.01^\circ$ ($68.55^\circ$) & $108.34^\circ$ ($27.28^\circ$) & 29.9\\
Sodankyl\"a (SOD) & $63.90^\circ$ ($67.42^\circ$) & $106.89^\circ$ ($26.39^\circ$) & 30.0\\
Rovaniemi (ROV) & $63.26^\circ$ ($66.78^\circ$) & $106.13^\circ$ ($25.94^\circ$) & 32.4\\
Oulu (OUL) & $61.51^\circ$ ($65.08^\circ$) & $105.15^\circ$ ($25.90^\circ$) & 30.0\\
Jyv\"askyl\"a (JYV) & $58.77^\circ$ ($62.42^\circ$) & $103.34^\circ$ ($25.28^\circ$) & 32.4\\
\hline
Canadian chain: & & &\\
Taloyoak (TAL) & $78.62^\circ$ ($69.54^\circ$) & $-30.04^\circ$ ($266.44^\circ$) & 30.0\\
Rankin Inlet (RAN) & $72.53^\circ$ ($62.82^\circ$) & $-24.84^\circ$ ($267.89^\circ$) & 30.0\\
Eskimo Point (ESK) & $70.80^\circ$ ($61.11^\circ$) & $-27.70^\circ$ ($265.95^\circ$) & 30.0\\
Fort Churchill (CHU) & $68.57^\circ$ ($58.76^\circ$) & $-27.27^\circ$ ($265.91^\circ$) & 30.0\\
Gillam (GIL) & $66.25^\circ$ ($56.38^\circ$) & $-27.73^\circ$ ($265.36^\circ$) & 30.0\\
Island Lake (ISL) & $63.82^\circ$ ($53.86^\circ$) & $-27.40^\circ$ ($265.34^\circ$) & 30.0\\
Pinawa (PIN) & $60.13^\circ$ ($50.20^\circ$) & $-29.00^\circ$ ($263.96^\circ$) & 30.0\\
\hline
\multicolumn{4}{l}{\textit{Note.} Locations are in geomagnetic coordinates with geodetic coordinates in brackets.}
\end{tabular}
\label{tab:stations}
\end{table}

The SGO and GO-Canada wide-beam riometers are analog La Jolla receivers with a dual half-wavelength dipole antenna that produces a single 60$^\circ$ beam towards the local zenith~\citep{spanswick2005}. The QDCs for the SGO riometers have been calculated with an automated method that fits a sinusoidal curve to data from the previous ten days to calculate the QDC for the next day. The QDC is calculated separately for each station. SGO data during winter months of 2000 to 2003 (27 events in total) were excluded from this study due to data being corrupted by unknown daily radio interference. The GO-Canada baselining method is based on characterizing the shape of the cosmic background noise rather than fitting a curve to a specific subset of data and is described in detail online at \texttt{http://aurora.phys.ucalgary.ca/GO-Canada/rio/doc/\linebreak CANOPUS\_Riometer\_Baselining.pdf}.
 Time resolution of the SGO data is one minute and the time resolution of the GO-Canada data is five seconds.

The Kilpisj\"arvi imaging riometer (IRIS) produces 49 narrow directional beams including a beam directed at the local zenith with a beam width of $13^\circ$~\citep{browne1995}. The QDC is produced separately for each beam by finding the largest value received without interference and absorption events for any point of time of a sidereal day. Observations from a time period of 14 days is usually used for the QDC determination. CNA measurements from the middle beam of the riometer was used in this study to produce a comparable beam to the modeled zenithal CNA. The time resolution of IRIS is one second, but the data were provided as five minute median values.

The imaging riometer in Longyearbyen produces 64 narrow directional beams. The QDC is determined for each beam separately by superimposing 10 to 20 days of observations near and including the day of interest into a mass plot, and determining the upper level of undisturbed observations visually~\citep{stauning1995}. As the riometer does not have a beam directed at the local zenith, CNA measurements from the four middle beams of the imaging riometer are combined into a single CNA value by taking the mean of the beam values if there are one or two measured values. If there are three or four values, the maximum and minimum values are discarded and the mean is taken from the remaining values. The time resolution of the Longyearbyen data is one minute.

Data from all riometers were averaged to have a five minute time resolution and checked manually. Times with clear abrupt level changes, QDC problems, and other clear radio interference were removed. The data was then averaged to one hour time resolution to match the time resolution of the WACCM-D model. The mean values of the standard error from averaging the data from five minute resolution into one hour resolution is less than 0.05 dB for all stations. The absorption measured by riometers not operating at 30 MHz was converted to 30 MHz equivalent absorption using the generalized magnetoionic theory $f^{-2}$ dependence of absorption and operating frequency~\citep{friedrich2002}. It should be noted, that the frequency dependence deviates from the inverse square relationship when strong spatial gradients of absorption regions are in the riometer beam and at altitudes below about 70 km altitude where the effective electron-neutral collision frequency becomes comparable with, or much greater than, the effective angular radio frequency~\citep{rosenberg1991}. Despite these caveats, the simple inverse square dependence was used, as the frequency dependence in the model CNA calculation method is close to the inverse square relationship.

Proton flux measurements during the studied SPEs are from the Space Environment Monitor (SEM) instrument package of the NOAA Geostationary Operational Environmental Satellite system (GOES) satellites. Due to the long time period of the study, data from two different GOES satellites had to be used. GOES-8 data was used for years 2000 to 2002, and GOES-10 data for years 2003 to 2005. Proton flux data from the $>10$ MeV integral proton flux channel (I3) was used to determine the durations of the SPEs.

\section{Modeling}
\label{sec:modeling}

WACCM-D is a variant of the global 3-D climate model, Whole Atmosphere Community Climate Model (WACCM), with added \textit{D}~region ion chemistry. The aim of the added \textit{D}~region chemistry is to better reproduce the effects of EPP on the neutral atmospheric constituents in the mesosphere and upper stratosphere. For a comprehensive description of WACCM-D and its lower ionospheric performance, see~\citet{verronen2016}. \citet{andersson2016} showed that the addition of \textit{D}~region ion chemistry into WACCM significantly improves modeling of polar HNO$_3$, HCl, ClO, OH, and NO$_x$, and that WACCM-D can model atmospheric effects of the January 2005 SPE (event 58 in this study, max 5040 pfu) as compared to Aura/MLS observations.

SD-WACCM-D (WACCM4) was run for the time periods of the 62 SPEs examined in this study with pre--configured specified dynamics driven by MERRA 19x2~\citep{rienecker2011} meteorological fields for the year 2000 with a six hour time resolution. The specified dynamics force the model at altitudes below 50 km at every dynamics time step by $10\%$, while model dynamics are fully interactive above 60 km. Between 50 km and 60 km altitude, the forcing transitions linearly from forcing to no forcing~\citep{kunz2011}. The model runs span the altitude range from the Earth's surface to the thermosphere ($4.5\cdot 10^{-6}$ hPa, $\approx 140$ km) with 88 vertical pressure levels. The latitudinal resolution of the model runs is $1.9^\circ$ and the longitudinal resolution is $2.5^\circ$. The specified dynamics driven SD-WACCM-D that was used in this study, is referred to as WACCM-D in the rest of this article.

Ionization sources in the used version of WACCM-D include solar protons, energetic radiation belt electrons, solar EUV, Lyman-$\alpha$, auroral electrons, and galactic cosmic rays. Hourly solar proton ionization rates were used in the model runs for the SPE protons. SPE ionization was applied uniformly to geomagnetic latitudes larger than $60^\circ$. The solar proton ionization rates are determined in the same way from GOES proton flux measurements as the daily ionization rates published by~\citet{jackman2005}, but with a higher time resolution. For an overview of the daily SPE ionization rate determination, see~\citet{jackman2013}. Ionization rates for energetic radiation belt electrons (30--1000 keV) were implemented from the medium-energy electron (MEE) model by~\citet{kamp2016}. The MEE model is based on precipitation data from low Earth orbiting POES satellites and an empirically described plasmasphere structure. The MEE model can use the \textit{Dst} or the \textit{Ap} index as an input and calculates the energy-flux spectrum of precipitating electrons with a time resolution of one day. In this study, we used the \textit{Ap}-driven model. The forcing from the MEE model includes electrons with energies from 30 to 1,000 keV which precipitate into 16 geomagnetic latitude bins between $45^\circ$ and $72^\circ$. Other ionization sources used in the model runs were standard WACCM ionization sources, i.e., solar EUV radiation, galactic cosmic rays (GCRs), auroral electrons, and solar Lyman-$\alpha$~\citep[see][]{marsh2007,smithjohnsen2018}. The time resolutions of the EUV and Lyman-$\alpha$ ionization sources are one day, and the time resolution of the auroral electron ionization source is three hours. The Nowcast of Atmospheric Ionising Radiation for Aviation Safety (NAIRAS) model is used in the simulation as the GCR ionization source~\citep{jackman2016}. The inclusion of GCR ionization in the model is necessary to provide an ion source for the \textit{D}~region chemistry at low latitudes.

The time resolution of the used WACCM-D output data is one hour or one month, depending on the atmospheric quantity. One hour resolution data is output by WACCM-D as a snapshot of the model state every hour at every model grid point and pressure level. Monthly data is output as a monthly mean of the wanted quantity at every model grid point and pressure level. The output data used in this study and their time resolutions are listed in Table~\ref{tab:dataProducts}.

\begin{table}
\caption{\textit{WACCM-D Output Data Used in This Study}}
\centering
\begin{tabular}{lc}
\hline
Data product & Time resolution\\
\hline
Neutral temperature (K) & Hourly \\
Electron mixing ratio, ppv & Hourly \\
O mixing ratio, ppv & Hourly \\
H mixing ratio, ppv & Hourly \\
O$_2$ mixing ratio, ppv & Monthly \\
N$_2$ mixing ratio, ppv & Monthly \\
\hline
\end{tabular}
\label{tab:dataProducts}
\end{table}

\begin{figure}[h]
 \centering
\centerline{\includegraphics[width=\textwidth]{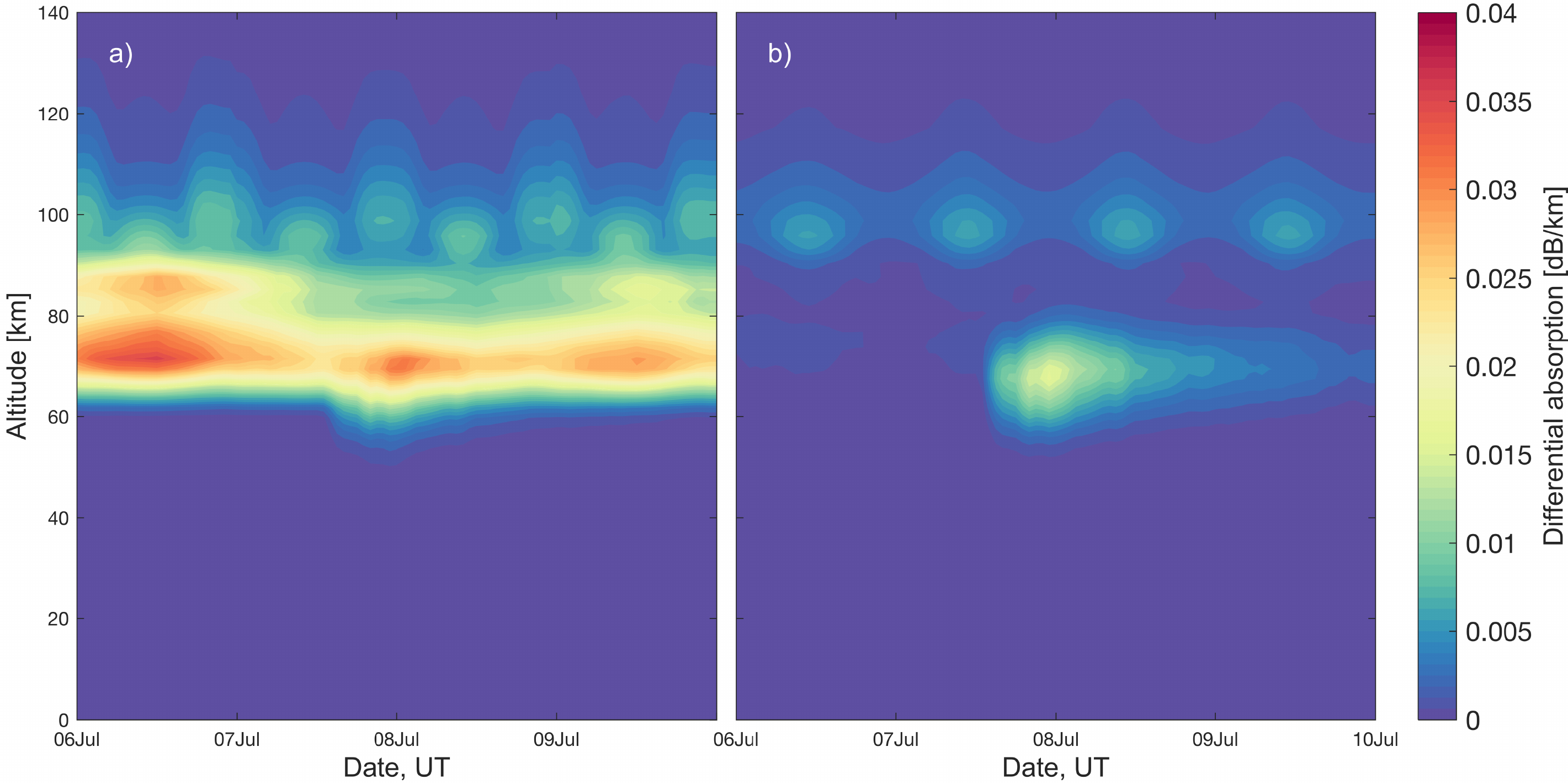}}
 \caption{Modeled differential CNA during the 7 July 2002 SPE (event 43, max 22 pfu) in Sodankyl\"a, panel a), and Longyearbyen, panel b). Date is is shown in the horizontal axis, altitude in the vertical axis, and differential CNA as a color-coded surface. Note that the plotted data starts from 00 UT 6 July 2002.}
 \label{fig:exampleDiffAbs}
 \end{figure}

To convert the atmospheric conditions in the WACCM-D model into CNA, differential CNA (dB/km) was calculated from WACCM-D output with the method by~\citet{sen1960}. The required electron collision frequencies with different neutral species (N$_2$, O$_2$, O, H) were calculated from WACCM-D data following~\citet[Part A, p. 194]{banks1973}. This approach has been previously used with Sodankyl\"a Ion and Neutral Chemistry model (SIC) data~\citep[e.g.,][]{verronen2006,clilverd2007}. WACCM-D does not provide electron temperature separately, thus it was assumed to be the same as the neutral temperature, which is a valid assumption below approximately 120 km altitude. Concentrations of He were not available from the model output; however the electron collision frequency with He is approximately five orders of magnitude smaller than that of the dominant species at 50 to 90 km altitude, based on our test calculations using He concentrations from the MSISE-E-90 model~\citep[for a MSIS description, see][]{hedin1991}. The differential absorption was integrated with respect to altitude to get the total absorption of the atmospheric column. Differential absorption as a function of altitude and time in SOD and LYR during the 7 July 2002 SPE (event 43, max 22 pfu) is shown in Figure~\ref{fig:exampleDiffAbs} as an example. Panel a) of the figure shows differential absorption in SOD and panel b) in LYR.  A weak event was chosen as an example to show the differential absorptions from multiple ionization sources, as the SPE ionization dominates in stronger events. At SOD (panel a)), CNA due to auroral electrons and EUV radiation is visible at altitudes above approximately 90 km. Auroral CNA is centered around approximately each midnight and EUV CNA is centered around approximately each midday. At LYR (panel b)), EUV is the dominant source of CNA above 90 km altitude. CNA due to radiation belt electrons is visible in the altitude range 60 to 90 km as the dominant source of CNA in SOD during this weak SPE event. CNA due to the SPE is clearly visible in LYR in the altitude range of 55 to 80 km starting abruptly from approximately midday of 7 July. The SPE is also visible in SOD, but not as clearly due to the CNA caused by radiation belt electrons. The ionization rates due to the SPE in LYR and SOD are identical as the model SPE input is applied uniformly to geomagnetic latitudes over $60^\circ$.

Model CNA for each riometer station was calculated separately at the closest grid point to the station's location with 30 MHz operating frequency. The wide-beam riometers' CNA were calculated as a wide-beam and the imaging riometers' CNA as zenithal CNA. For the wide-beam riometers, the modeled zenithal CNA were multiplied by a scaling factor of 1.2 in the calculations to get equivalent wide-beam CNA~\citep{hargreaves1987}.

\section{Results}
\label{sec:results}

\begin{figure}[hp]
 \centering
\centerline{\includegraphics[width=\textwidth]{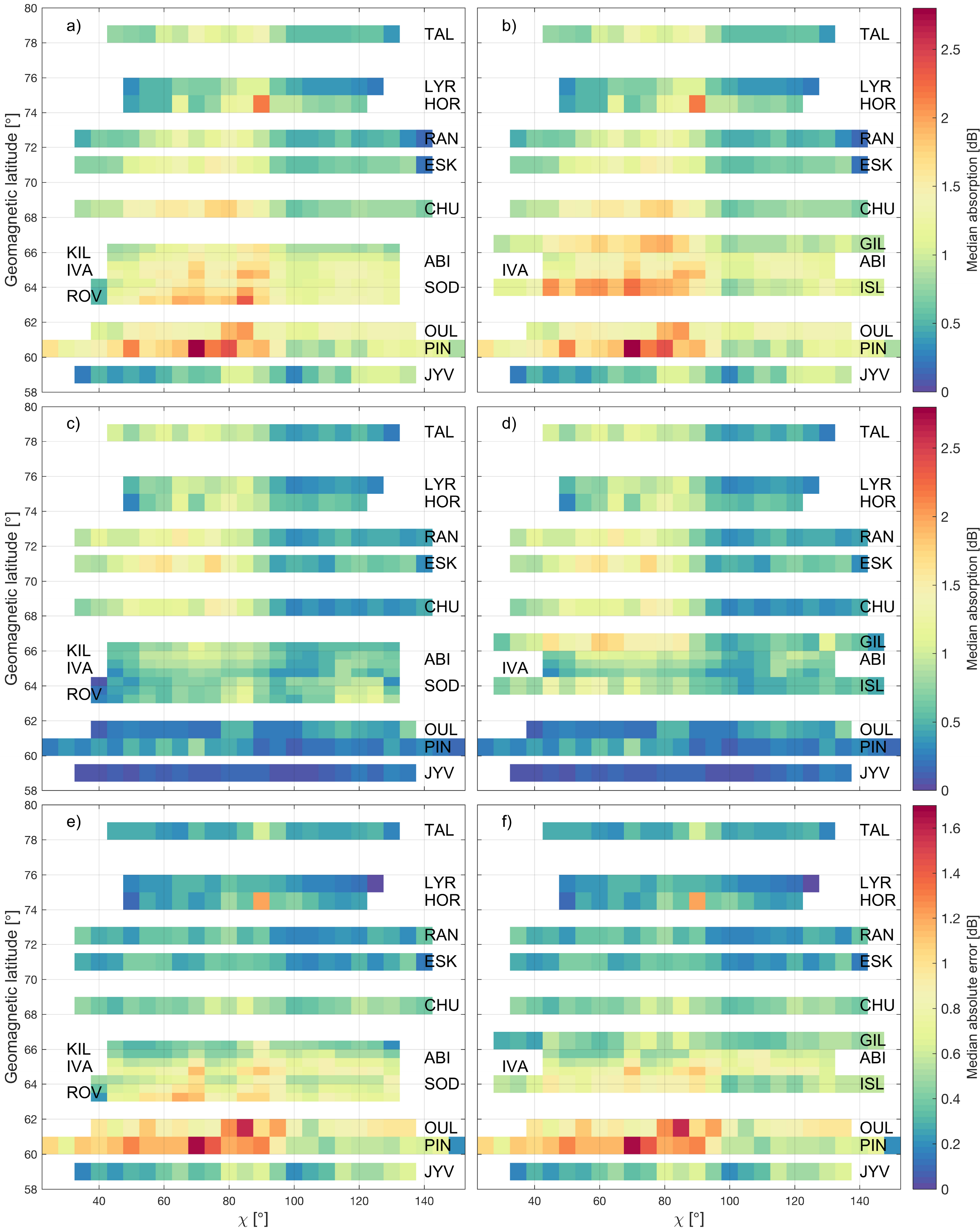}}
\caption{Modeled and observed median absorptions during SPEs, and the median absolute errors between the model and the observations as a function of solar zenith angle, $\chi$, and geomagnetic latitude. Panels in the left column are for case 1 (overlapping GO-Canada stations removed) and panels in the right column are for case 2 (overlapping European sector stations removed). Panels a) and b) are modeled median absorptions, panels c) and d) are observed median absorptions, and panels e) and f) are the median absolute errors between the model and the observations. Note that the color scaling in the last row is different from the first two rows.}
 \label{fig:zenithAngle}
 \end{figure}

The median observed and modeled CNA during the 62 SPEs were compared as a function of solar zenith angle and geomagnetic latitude. The field-of-view (FoV) of each riometer in the \textit{D}~region was assumed to be $\pm 0.5^\circ$ in geomagnetic latitude for the comparison, which corresponds approximately to the area seen by a $60^\circ$ wide-beam riometer at 90 km altitude (calculated to be $\pm 0.46^\circ$ geomagnetic latitude). The observational and model CNA data from all available SPEs were binned into $5^\circ$ solar zenith angle bins for each station and the median value of the bin was calculated. Data were limited to time periods where the GOES I3 integral proton flux is greater than or equal to 10 pfu and observational data is available. In case of overlapping riometer FoVs, the overlapping bins were averaged together. At any latitude bin with both GO-Canada and European chain riometer data, only one chain is used in the median value calculations by simply removing the other, which then favors either the European chain (case 1) or GO-Canada (case 2). In case 1, GIL and ISL were removed, and in case 2, KIL, SOD, and ROV were removed. This approach was chosen due to the difference in observed CNA between the the two riometer chains, which is due to the removal of the corrupted winter events from the SGO data and the different QDC methods. Model median absorption and observed median absorption for case 1 during the SPEs as a function of geomagnetic latitude and solar zenith angle are shown in panels a) and c) of Figure~\ref{fig:zenithAngle}. Model median absorption and observed median absorption for case 2 during the SPEs as a function of geomagnetic latitude and solar zenith angle are shown in panels b) and d) of Figure~\ref{fig:zenithAngle}, and the median absolute errors between the modeled and observed absorptions for both cases are shown in panels e) and f), respectively. Bins with less than or equal to ten data points were removed in each of the panels. The number of data points in the remaining bins varies between 15 and 268 from the extreme solar zenith angles to the most common solar zenith angles.

The median absolute errors between the model and the observations are very similar at the five stations (TAL--ESK) poleward of $70^\circ$ geomagnetic latitude. The model underestimates the CNA slightly as compared to the observations in the sunlit atmosphere and overestimates it in the twilight transition. CNA in the dark atmosphere is overestimated slightly by the model as compared to the observations. The differences and median absolute errors are generally small ($\leq 0.5$ dB) poleward of $70^\circ$ geomagnetic latitude. The mean values of the differences between the model and the observations during sunlit ($\chi < 82.5^\circ$), twilight ($ 82.5^\circ < \chi < 97.5^\circ$), and dark conditions ($\chi > 97.5^\circ$) for all 16 stations are listed in Table~\ref{tab:stationParams}. The absorption decrease due to the twilight transition is at larger zenith angles in the model than in the observations, which can be seen as increased difference between the model and the observations in twilight conditions and as increased median absolute errors in the zenith angle bin centered at $\chi = 90^\circ$ in panels e) and f) of Figure~\ref{fig:zenithAngle}.
The two GO-Canada stations between $66^\circ$ and $69^\circ$ geomagnetic latitude (CHU and GIL) show larger differences and median absolute errors between the model and the observations than the poleward stations. Unlike at the stations poleward of $70^\circ$ geomagnetic latitude, the sunlit values are generally overestimated by the model. CNA in the dark atmosphere and during twilight conditions is overestimated by the model at these two stations.
The difference between the model and the observations at the European chain stations KIL--ROV is systematically larger than the difference at the poleward stations and GIL. The overestimation of CNA by the model compared to the observations increases with decreasing geomagnetic latitude.
The results from the stations equatorward of approximately $66^\circ$ geomagnetic latitude indicate that protons precipitating into these geomagnetic latitudes are subject to varying levels of geomagnetic cutoff, which is not represented in the simulations, and that the MEE ionization is overestimated in the model. The effect of geomagnetic cutoff is especially evident at OUL and PIN, where CNA is overestimated by the model at all zenith angles. The overestimation of MEE ionization is, in conjunction with the geomagnetic cutoff effect, responsible for the overestimation of CNA at geomagnetic latitudes between about $63^\circ$ and $66^\circ$. The modeled and observed median absorptions in JYV are low, as the latitude limit for proton precipitation in the model is set at $60^\circ$ geomagnetic latitude and only very high energy (about $ > 100$ MeV~\citep[Figure 8]{rodger2006}) protons can precipitate into the atmosphere above JYV in the observed data.

\begin{table}
\caption{\textit{Mean Differences Between the Model and the Observed Median CNA at Each Riometer Station, and the Non-Linearity Correction Parameter $R$ and Its $95\%$ Confidence Intervals}}
\centering
\begin{tabular}{lccccc}
\hline
Station & Sunlit$^{a}$ (dB) & Twilight$^{b}$ (dB) & Dark$^{c}$ (dB) & $R$ & $95\%$ CI\\
\hline
TAL & 0.06 & 0.45 & 0.18 & 4.02 & 3.69/4.40 \\
LYR & -0.24 & 0.14 & 0.07 & & \\
HOR & 0.06 & 0.62 & 0.29 & 1.88 & 1.72/2.05 \\
RAN & -0.16 & 0.22 & 0.09 & 11.93 & 9.51/14.85 \\
ESK & -0.16 & 0.24 & 0.14 & 17.84 & 14.39/21.67 \\
CHU & 0.25 & 0.50 & 0.43 & 4.19 & 3.53/4.97 \\
GIL & 0.23 & 0.47 & 0.38 & 4.25 & 3.65/4.95 \\
KIL & 0.29 & 0.59 & 0.36 & & \\
ABI & 0.34 & 0.59 & 0.49 & 7.77 & 4.76/18.16 \\
IVA & 0.78 & 1.00 & 0.76 & 1.98 & 1.72/2.31 \\
ISL & 0.75 & 0.70 & 0.44 & 3.71 & 3.02/4.65 \\
SOD & 0.63 & 0.74 & 0.52 & 2.00 & 1.74/2.35 \\
ROV & 0.79 & 0.85 & 0.59 & 9.88 & 5.61/21.02 \\
OUL & 0.93 & 1.25 & 0.82 & & \\
PIN & 1.06 & 1.02 & 0.45 & & \\
JYV & 0.31 & 0.41 & 0.39 & & \\
\hline
\multicolumn{6}{l}{$^{a}$ Sunlit $\chi < 82.5^\circ$. Twilight $^{b}$ $82.5^\circ < \chi < 97.5^\circ$. $^{c}$ Dark $\chi > 97.5^\circ$.}
\end{tabular}
\label{tab:stationParams}
\end{table}

\begin{figure}[h]
 \centering
\centerline{\includegraphics[width=\textwidth]{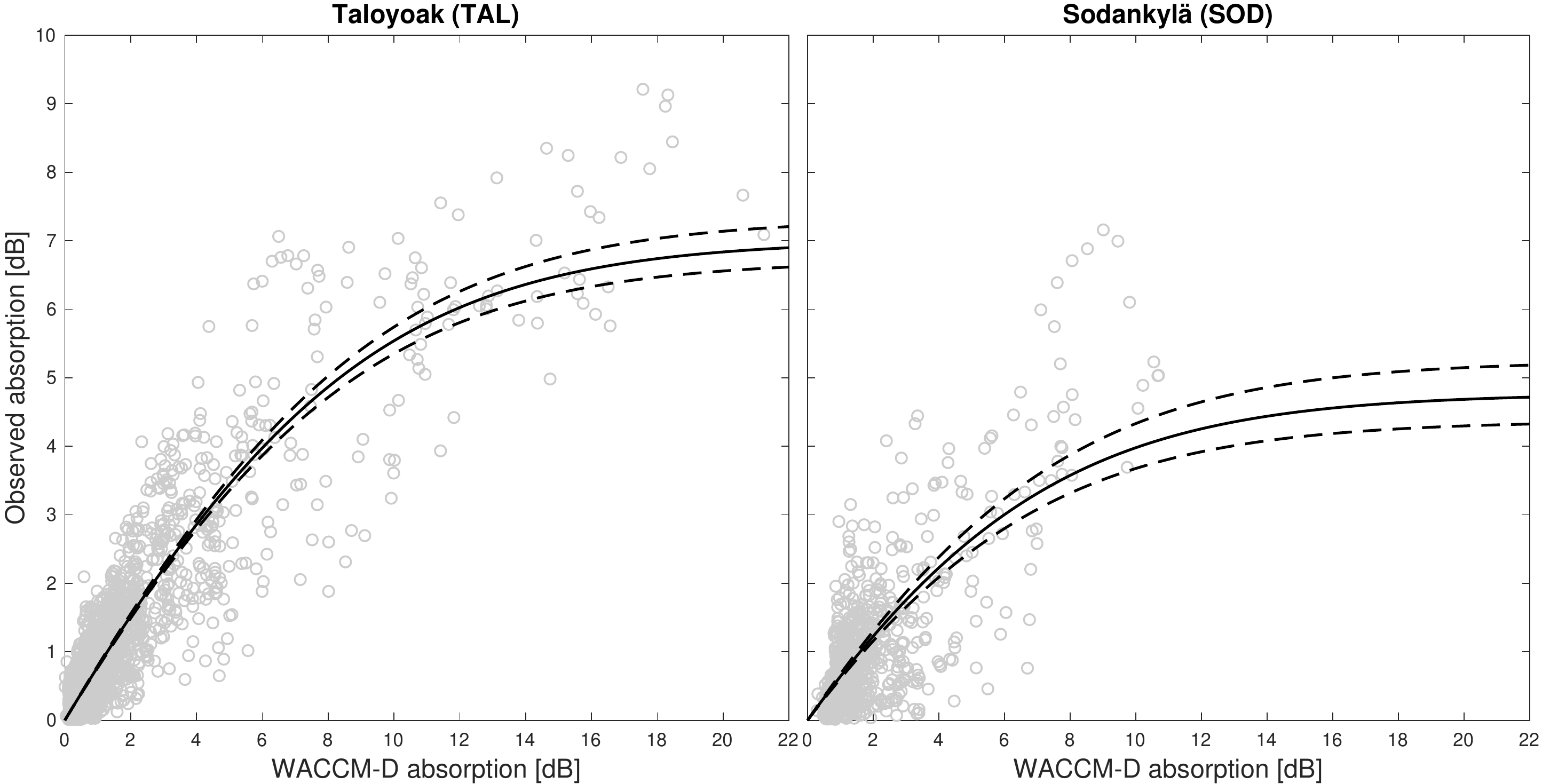}}
 \caption{Modeled and observed absorptions from TAL, panel a), and SOD, panel b), riometers with fitted non-linear response curves, Equation~(\ref{eq:fitFunc}), and their $95\%$ bootstrap confidence intervals. Individual data points shown as gray circles, best fits to data as solid lines, and confidence intervals as dashed lines.}
 \label{fig:scatter}
 \end{figure}

The relationship between the observed and modeled CNA during the SPEs was studied by plotting the one hour data points during the SPEs as scatter plots for each station. Scatter plots for TAL (panel a)) and SOD (panel b)) are shown as an example in Figure~\ref{fig:scatter}. The observed and modeled CNA values at TAL agree well and are linear to approximately 6 dB, with an intercept at the origin. At higher model CNA values, the relationship between the observed and modeled CNA becomes non-linear and the model shows higher values than the observations. The TAL scatter plot was chosen as an example of best-case non-linear agreement between the observed and modeled CNA of the wide-beam riometers. Compared to the TAL scatter plot, the SOD scatter plot shows a worst-case agreement between observed and modeled values. The modeled CNA values are offset from zero by approximately 0.5 dB when the observed CNA values are at zero. A similar offset is visible in IVA, and to a lesser extent in ROV, GIL, ABI, and HOR (not shown). Compared to TAL, the observed CNA values at SOD vary more with modeled values between 1 and 2 dB, and the linear and non-linear relationships are more difficult to discern. Increased variation in observed CNA values at modeled values of approximately 1 to 2 dB is also visible in ROV, IVA, ISL, GIL, and ABI. The observed absorptions from LYR and KIL imaging riometers are linear with model absorption (not shown) with separate populations for sunlit and dark atmospheric conditions. The slopes for LYR and KIL in sunlit conditions are 1.30 and 1.00, and in dark conditions 0.71 and 0.61.

As the southernmost stations are heavily affected by geomagnetic cutoffs and the simulations do not consider the effect of geomagnetic cutoff poleward of $60^\circ$ geomagnetic latitude, the data from JYV, PIN, and OUL were removed from the comparison. As shown in the example scatter plots of Figure~\ref{fig:scatter}, the response of the wide-beam riometers becomes non-linear with large modeled absorption values. All analyzed wide-beam riometers have a non-linear response to large modeled CNA, but the dependence between the observed and modeled values varies from station to station. A possible correction method for the non-linear response is presented here.

During normal conditions of low ionospheric absorption the noise power available to a riometer is determined by the radio noise temperature of the sky with negligible contributions from the absorbing region of the ionosphere and losses in the receiving instrumentation~\citep{little1958}. As the absorption of the ionosphere increases to large values (greater than 10 dB~\citep{browne1995}), these normally negligible effects become significant. At large absorption values, the riometer receives additional significant signals from the absorbing ionosphere~\citep{hargreaves2002} and the lossy hardware causing the riometer response to become non-linear. The signal, $P$, measured by a riometer is:
\begin{linenomath*}
\begin{equation}
P = G (a \cdot T_\textrm{s} + T_\textrm{r}),
\end{equation}
\end{linenomath*}
where $G$ is the gain of the instrument, $a$ is absorption as a linear value in range $[0,1]$, $T_\textrm{s}$ is the wanted sky noise measured by the instrument, and $T_\textrm{r}$ is unwanted noise from other sources. Absorption, $A$, is given in data as decibels compared to a quiet day marked by subscript $\textrm{q}$. Absorption is therefore given by
\begin{linenomath*}
\begin{equation}
A = 10\cdot \log_{10} \left( \frac{P_\textrm{q}}{P} \right) = 10\cdot \log_{10} \left( \frac{a_\textrm{q} \cdot T_\textrm{s} + T_\textrm{r}}{a \cdot T_\textrm{s} + T_\textrm{r}} \right).
\end{equation}
\end{linenomath*}
The ratio between the wanted and unwanted noise is
\begin{linenomath*}
\begin{equation}
R = \frac{T_\textrm{s}}{T_\textrm{r}}.
\end{equation}
\end{linenomath*}
Assuming that the quiet day absorption is small $(a_\textrm{q} = 1)$:
\begin{linenomath*}
\begin{equation}
A = 10\cdot \log_{10} \left( \frac{1 + 1/R}{a + 1/R}  \right),
\end{equation}
\end{linenomath*}
where $a = 10^{-A_\textrm{s} /10}$ and $A_\textrm{s}$ is the true absorption of the ionosphere in dB. As $R$ increases, the absorption, $A$, approaches the true absorption of the ionosphere $A_\textrm{s}$:
\[ R \rightarrow \infty \Rightarrow A \rightarrow A_\textrm{s} \]
The dependence between observed absorption and the true absorption of the ionosphere can be written as
\begin{linenomath*}
\begin{equation}
A = 10\cdot \log_{10} \left( \frac{1 + 1/R}{10^{-A_\textrm{s} /10} + 1/R}  \right). \label{eq:fitFunc}
\end{equation}
\end{linenomath*}
Assuming that the modeled WACCM-D absorption is the true absorption of the ionosphere, the presented dependence can be used to determine the ratio between wanted and unwanted noise in the riometer, which can be used to convert between observed absorption values and true absorption values. The correction function, Equation~(\ref{eq:fitFunc}), was fitted to each station separately with WACCM-D absorption as the true absorption and $R$ as a free parameter. A non-linear least squares method was used in the fitting. The fitted function for TAL and SOD, and the $95\%$ bootstrap confidence intervals, are shown in the scatter plots of Figure~\ref{fig:scatter}. $10,000$ bootstrap samples were used for the confidence interval determination. The $R$ values and their $95\%$ bootstrap confidence intervals for each analyzed wide-beam riometer are listed in Table~\ref{tab:stationParams}.

\begin{figure}[hp]
 \centering
\centerline{\includegraphics[width=\textwidth]{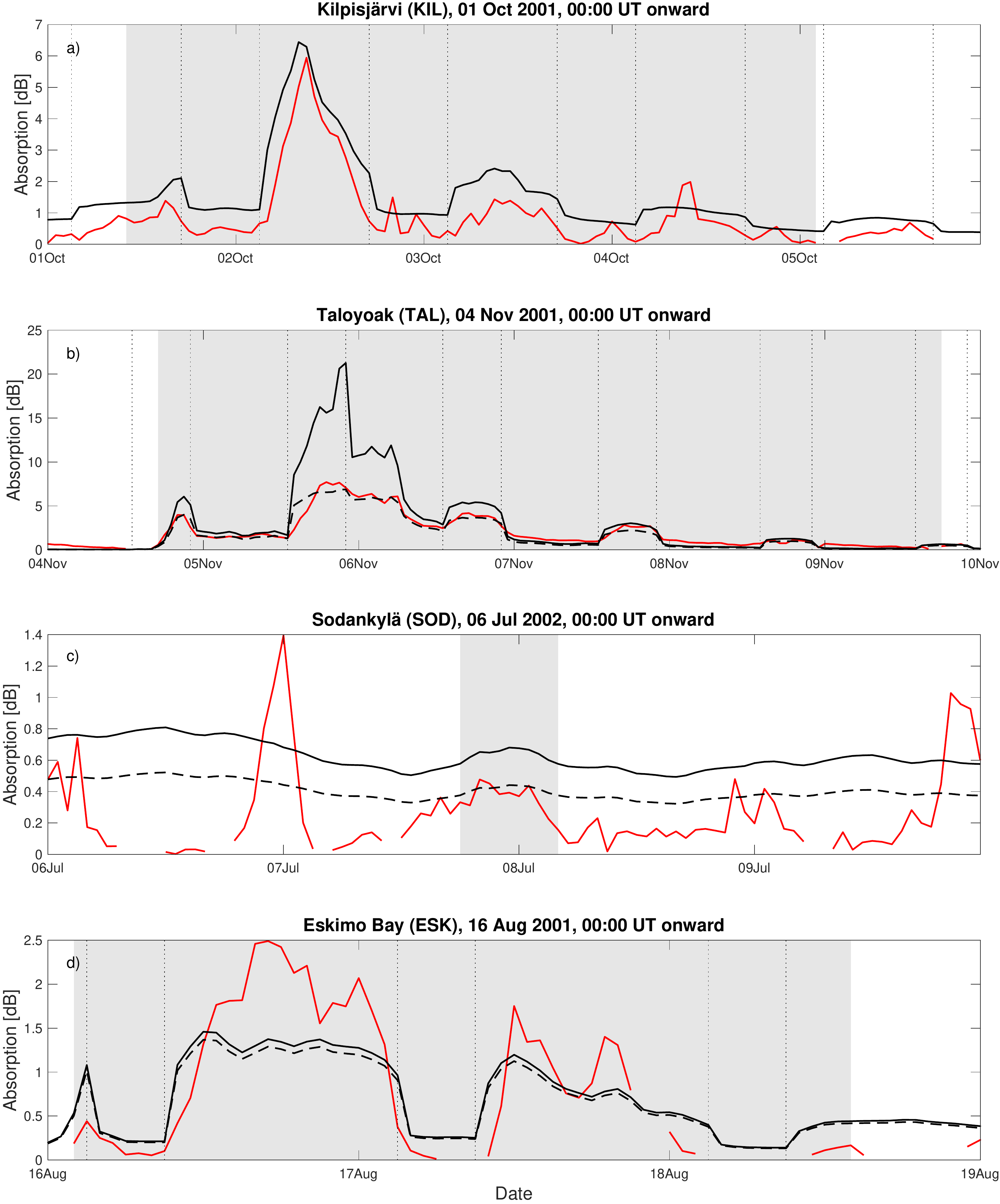}}
\caption{Examples of modeled and observed CNA during four SPEs at four different stations. Solid black lines are modeled CNA values, dashed black lines are modeled CNA values with the non-linear response correction applied, and the solid red lines are observations. Note, that panel a) does not have a non-linear response corrected line (black dashed line). The gray shaded areas indicate time periods where the GOES $>$ 10 MeV integral flux is greater than or equal to 10 pfu and the dotted vertical lines are the approximate times where the solar zenith angle is $97^\circ$.}
 \label{fig:events}
 \end{figure}

Modeled and observed absorptions were plotted as time series, separately for each station and event, to study the performance of WACCM-D in reproducing the temporal evolution of ionospheric absorption in individual events. Four individual events are shown as examples in Figure~\ref{fig:events}. Solid black lines are modeled CNA values, dashed black lines are modeled CNA values with the non-linear response correction applied, and the solid red lines are observations. The time periods where the GOES $>10$ MeV integral flux is greater than or equal to 10 pfu have been shaded and the approximate times of the model twilight change ($\chi = 97^\circ$) are marked with dotted vertical lines.
Panel a) shows event 26 (max $2,360$ pfu) from KIL, where the magnitude of the CNA is overestimated throughout the event by the model up to approximately 1 dB, but the time behavior corresponds very well with the observations. Note that the non-linear response correction is not applied in the top panel, as the response was linear for both of the imaging riometers (KIL and LYR).
Panel b) shows an extreme event (event 28, max $31,700$ pfu) from TAL with the best-case non-linear response correction. The modeled CNA reaches a maximum value of approximately 21 dB with an observed maximum value of approximately 7.5 dB. The modeled CNA corrected for the non-linear response of the riometer agrees very well with the observed CNA. The observed sunrise increase in CNA is more gradual than the abrupt increase in the modeled CNA, especially on 5 November.
A weak event (event 43, max 22 pfu) with auroral activity from the SOD riometer is shown in panel c). The same event at SOD is shown in Figure~\ref{fig:exampleDiffAbs} panel a). A small increase in CNA is seen during the short SPE in both observations and model data. Outside the SPE, the observed CNA peaks around midnight between 6 July and 7 July, and the evening of 9 July, are caused by auroral activity. The auroral activity is not well reproduced as the model ionization for auroral electrons (\textit{Kp} parametrization) produces a uniform ionization band at the auroral oval latitudes which cannot capture local variations (e.g., substorm activity) properly. Unlike the SPE ionization input, the auroral electron parametrization is based on magnetic field variations rather than direct particle measurements. The higher than observed CNA in the model due to the radiation belt electron input is visible through the plot.
In 18 of the studied events, the observed CNA was found to be higher than the modeled CNA in three or more stations during sunlit conditions. These cases are only present at stations poleward of $66^\circ$ geomagnetic latitude. One such event (event 24, max 493 pfu) from ESK is shown in panel d) of Figure~\ref{fig:events}. In this example event, CNA in sunlit conditions is underestimated by the model up to approximately $1.1$ dB and CNA in dark conditions is overestimated by the model up to 0.25 dB. As in panel b), the sunrise increase in CNA is more abrupt in the model than in the observations. In addition to the model sunrise increase of CNA being more abrupt, the sunset decrease in CNA is also more abrupt in the model than in the observations. Variations in the model CNA for the example events of Figure~\ref{fig:events} were estimated by calculating the mean absolute error of the CNA at the station's grid point and the adjacent grid points (nine grid points in total) at each time step. The mean absolute errors were less than 0.08 dB at all time time steps discarding the points closest to the model's twilight transition. Next to the twilight transition, where the adjacent grid point is on the other side of the twilight transition, the maximum error in the example events was 0.95 dB in the extreme event shown in panel b) of Figure~\ref{fig:events}.

\section{Discussion}
\label{sec:discussion}

WACCM-D can model CNA well in the polar cap, both in sunlit and dark conditions. The twilight transition of CNA is not modeled as well as the sunlit or dark conditions. The modeled median CNA values at twilight conditions are higher than in the observations, which indicates that the increase in CNA during sunrise is more abrupt in the model than in observations and that the sunset decrease of CNA is delayed compared to observations. These differences in sunrise and sunset behavior are also visible in panels a), b), and d) of Figure~\ref{fig:events}. The twilight difference between the model and the observations is likely due to the night/day scaling in WACCM-D, which is a simple on/off at the solar terminator in the \textit{D}~region at $\chi = 97^\circ$~\citep{verronen2016}, i.e, an Earth shadow method. In reality, electron depletion starts during sunset when the whole mesosphere is still sunlit~\citep{collis1990} and the rise in CNA can be delayed at sunrise by a screening of solar UV radiation by the stratospheric ozone layer and the slowly developing chemical changes in the \textit{D}~region~\citep[for a review, see][and references therein]{rogers2016}. Similar twilight difference results were presented by~\citet{rogers2016}, who concluded that using the Earth shadow method in full-profile CNA models will fail to represent the slowly varying ionospheric composition and temperature changes affecting CNA at presunset and especially postsunrise conditions.

Cutoff latitudes are not static, but move in latitude with changing conditions of the magnetosphere and the solar wind from event to event, and even within events~\citep{nesse-tyssoy2015}, producing a gradual effect in the averaged data. The gradual effect of the geomagnetic cutoffs on observed CNA is visible in Figure~\ref{fig:zenithAngle} and Table~\ref{tab:stationParams} at stations equatorward of approximately $66^\circ$ geomagnetic latitude. The effect is strongest in the stations between $60^\circ$ and $62^\circ$ (OUL and PIN), where the energetic protons responsible for the majority of the CNA are almost completely cutoff in the observations on average. 
The decrease in observed median CNA, increasing differences between the model and the observations, and increasing median absolute errors, especially during sunlit conditions, with decreasing geomagnetic latitude between $66^\circ$ and $63^\circ$ indicate that increasing amounts of protons responsible for CNA are cutoff in the averaged data. During suitable conditions, lower energy protons can access lower latitudes and push the average main cutoff effect equatorward. When limiting data to events with a maximum flux greater than $1,000$ pfu, the main cutoff effect in the averaged data seems to be pushed to about $61^\circ$ geomagnetic latitude. This agrees with the results by~\citet{rodger2006} in that the larger geomagnetic disturbances associated with large SPEs increase the impact area of SPE particles. In addition to moving in latitude with varying magnetospheric and solar wind conditions, cutoff latitudes have been shown to have day-night asymmetry~\citep{nesse-tyssoy2013,nesse-tyssoy2015}, which is not taken into account by the static cutoff latitude of the model.

The commonly used~\citep[e.g.,][]{jackman2009, matthes2017} $60^\circ$ lower geomagnetic latitude limit for uniform SPE proton precipitation and ionization overestimates the spatial extent of the SPE effect. The importance of implementing improved geomagnetic cutoff constraints for solar proton precipitation in chemistry climate models depends on the desired accuracy of the model results, the timescale of the model studies, and the aggregated impact of SPEs on the chemistry of the middle atmosphere compared to other ionization sources. Even though SPEs cause large-scale ionization and chemical changes in the middle atmosphere, they occur rarely compared to the precipitation of high-energy electrons from the radiation belts during geomagnetically active periods or during substorms. The chemical effect of radiation belt electrons during a single large geomagnetic storm on the neutral atmosphere was modeled by~\citet{rodger2010} with the SIC model. They reported mesospheric O$_3$ changes that are fairly similar in magnitude, time scales, and altitude to those presented in previous model and experimental observation studies for large solar proton events. \citet{seppala2015} modeled a five day period with 61 substorms using the SIC model resulting in mesospheric O$_3$ changes that are similar in scale to a small to medium SPE. Compared to energetic electron precipitation (EEP) effects on the neutral atmosphere, SPEs affect the whole polar cap area and ionize the atmosphere to lower altitudes. At least some of the outer radiation belt electrons precipitate outside the polar vortex, leading to less NO$_x$ transport during the polar winter than in SPEs and making the direct contrasting of EEP and SPE effects more difficult~\citep{rodger2010}.

The modeled CNA is overestimated compared to observations in the auroral and subauroral latitudes as seen in Figure~\ref{fig:zenithAngle} and panel b) of Figure~\ref{fig:scatter}. The overestimation is in part due to the $60^\circ$ cutoff latitude and in part due to the overestimation of MEE ionization. The used MEE model does not have an MLT dependence, so a daily zonal mean MEE flux is applied uniformly to all MLTs. The use of daily zonal mean forcing results in overestimation of MEE fluxes on the dayside and underestimation on the nightside. Like the auroral electron ionization input, the MEE model is statistical and is not able to reproduce local variations in MEE precipitation. The effective recombination coefficient is smaller in the sunlit \textit{D}~region than in the dark \textit{D}~region~\citep{hargreaves2005} resulting in larger CNA in sunlit conditions for the same ionization forcing. Therefore overestimation in the MEE forcing will cause a larger increase in model CNA in the sunlit atmosphere than in the dark atmosphere, as seen from the model results. An additional reason for the overestimation of CNA by the model in the sunlit \textit{D}~region can be solar radio emission (SRE), which will cause reduced observed absorption~\citep{kavanagh2004}, especially when the Sun is in the riometer beam or in the beam side lobes. The WACCM-D model also overestimates CNA in the dark atmosphere at auroral and subauroral latitudes, but the differences between the model and the observations are smaller. The used MEE model has been recently refined by~\citet{kamp2018} with improved consideration for low electron fluxes and an option to include MLT-dependence with three-hour temporal resolution. The performance of WACCM-D with the refined MEE model with MLT-dependence should be investigated in the future.

The non-linear response of riometers to large levels of absorption is a known phenomenon in the field, but rarely discussed in publications. The hardware in riometer systems have been designed and configured in different ways resulting in different upper limits for the linear response. The non-linear response correction presented in this paper works well for some of the used riometers, but is sensitive to data selection. Special care has to be taken during the selection of data for the method, as the uncertainties in fitting of the $R$ parameter can become large as shown by the $95\%$ bootstrap confidence intervals listed in Table~\ref{tab:stationParams}. The poor fit in some of the stations is due to the variation in observed CNA values due to substorms, geomagnetic storms, and geomagnetic cutoff, variation in modeled CNA values due to the MEE model's zonal mean input, and the different atmospheric conditions in the sunlit and dark atmosphere.  Based on a cursory examination, and the fact that the largest CNA values occur in the sunlit atmosphere, data should possibly be limited to sunlit conditions when determining the non-linearity of a riometer with this method. When data are limited to sunlit conditions, the values of the $R$ parameter become close to or higher than the upper limits of the $95\%$ bootstrap confidence intervals. Further testing of the non-linear correction method falls outside the scope of this paper, but the presented results should provide a good starting point for future work on the subject.

No explanation was found for the situations where the dayside observed CNA is higher than the modeled CNA in multiple stations (see panel d) of Figure~\ref{fig:events}). The higher observed CNA indicates that either the model is underestimating some ionization source or that some other phenomenon causing ionization is lacking from the model in these cases. The difference in CNA between the model and the observations was contrasted with geomagnetic activity (\textit{Kp} index), integral flux in the different GOES proton channels, season, and cases where multiple SPEs occur consecutively, but no explanation was found for the underestimated model CNA. The electron density output of the model was increased as a test for the example event and location shown in panel d) of Figure~\ref{fig:events}. An increase in electron density by a factor of 1.5 to 1.75 produced approximately the correct level of CNA, but did not affect the time behavior of the model. As the stations where this underestimation occurs are poleward of about $66^\circ$ geomagnetic latitude, a possible explanation is that the SPE proton precipitation is underestimated in the model. This would not explain however why the underestimation is not visible in all or most events. It is also possible that the underestimation happens at all latitudes in the model, but that the MEE precipitation and geomagnetic cutoff effects mask it from stations equatorward of about $66^\circ$ geomagnetic latitude.

\section{Conclusions}
\label{sec:conclusions}

We have studied the spatial and temporal extent of CNA during 62 SPEs from 2000 to 2005 using the WACCM-D model and observations from 16 riometer stations. Observed and modeled CNA were contrasted as a function of solar zenith angle and geomagnetic latitude statistically, for each station statistically, and as time series for each event and station individually. We summarize the results of this study as follows:
\begin{enumerate}
\item WACCM-D can reproduce the observed CNA well poleward of about $66^\circ$ geomagnetic latitude with an average absolute difference between the model and the observations of less than $0.5$ dB varying with solar zenith angle and station.
\item Equatorward of approximately $66^\circ$ geomagnetic latitude, the average difference between the model and the observation increases with decreasing geomagnetic latitude from about $0.5$ dB to $1$ dB due to the daily zonal mean MEE forcing and the uniform proton forcing poleward of $60^\circ$ geomagnetic latitude. 
\item Due to the Earth shadow implementation of the change between night and day in WACCM-D, the CNA increase (decrease) during sunrise (sunset) is more abrupt and at greater solar zenith angle values in WACCM-D than what is observed, resulting in overestimation of CNA during twilight conditions.
\item Observed CNA in sunlit conditions is underestimated  by WACCM-D at three or more stations poleward of $66^\circ$ geomagnetic latitude in 18 events, in contrast with WACCM-D usually overestimating the observed CNA. More investigations are required to explain the underestimation of CNA in sunlit conditions by WACCM-D for this subset of events.
\item The absorption response of the used wide-beam riometers becomes non-linear at large absorption values. A correction method for this non-linearity was presented with the goal of providing a starting point for further studies on the subject.
\item The used $60^\circ$ stationary cutoff latitude for proton precipitation in WACCM-D was found to overestimate the spatial extent of CNA during SPEs by about $2^\circ$ to $3^\circ$ geomagnetic latitude on average. The overestimation of the average spatial extent seems to decrease to about $1^\circ$ geomagnetic latitude when data are limited to events with a maximum flux greater than $1,000$ pfu. A more realistic cutoff model should be implemented into the proton precipitation forcing in the future, if more accurate performance is required from the model. 
\end{enumerate}

Although the overall performance of WACCM-D in reproducing CNA is good, some tests and improvements are recommended for the future.  The results presented here should be compared with comparison runs of the WACCM-D model with improved proton cutoff constraints and MLT-dependent MEE-fluxes.

\appendix
\section{List of studied solar proton events}
\label{sec:appendix}

\begin{table}
\caption{\textit{Solar Proton Events Used in This Study}}
\centering
\begin{tabular}{lccc}
\hline
Event & SPE start time & SPE max time & Max $>10$ MeV flux \\ 
 & (UT) & (UT) & (pfu) \\ 
\hline
1$^{*}$ & 18 Feb 2000 11:30 & 18 Feb 12:15 & 13\\
2 & 04 Apr 2000 20:55 & 05 Apr 09:30 & 55\\
3 & 07 Jun 2000 13:35 & 08 Jun 09:40 & 55\\
4 & 10 Jun 2000 18:05 & 10 Jun 20:45 & 46\\
5 & 14 Jul 2000 10:45 & 15 Jul 12:30 & 24,000\\
6 & 22 Jul 2000 13:20 & 22 Jul 14:05 & 17\\
7 & 28 Jul 2000 10:50 & 28 Jul 11:30 & 18\\
8 & 11 Aug 2000 16:50 & 11 Aug 16:55 & 17\\
9 & 12 Sep 2000 15:55 & 13 Sep 03:40 & 320\\
10$^{*}$ & 16 Oct 2000 11:25 & 16 Oct 18:40 & 15\\
11$^{*}$ & 26 Oct 2000 00:40 & 26 Oct 03:40 & 15\\
12$^{*}$ & 08 Nov 2000 23:50 & 09 Nov 16:00 & 14,800\\
13$^{*}$ & 24 Nov 2000 15:20 & 26 Nov 20:30 & 942\\
14$^{*}$ & 28 Jan 2001 20:25 & 29 Jan 06:55 & 49\\
15$^{*}$ & 29 Mar 2001 16:35 & 30 Mar 06:10 & 35\\
16$^{*}$ & 02 Apr 2001 23:40 & 03 Apr 07:45 & 1,110\\
17$^{*}$ & 10 Apr 2001 08:50 & 11 Apr 20:55 & 355\\
18$^{*}$ & 15 Apr 2001 14:10 & 15 Apr 19:20 & 951\\
19$^{*}$ & 18 Apr 2001 03:15 & 18 Apr 10:45 & 321\\
20$^{*}$ & 28 Apr 2001 04:30 & 28 Apr 05:00 & 57\\
21 & 07 May 2001 19:15 & 08 May 07:55 & 30\\
22 & 15 Jun 2001 17:50 & 16 Jun 00:05 & 26\\
23 & 10 Aug 2001 10:20 & 10 Aug 11:55 & 17\\
24 & 16 Aug 2001 01:35 & 16 Aug 03:55 & 493\\
25 & 24 Sep 2001 12:15 & 25 Sep 22:35 & 12,900\\
26 & 01 Oct 2001 11:45 & 02 Oct 08:10 & 2,360\\
27$^{*}$ & 19 Oct 2001 22:25 & 19 Oct 22:35 & 11\\
28$^{*}$ & 04 Nov 2001 17:05 & 06 Nov 02:15 & 31,700\\
29$^{*}$ & 19 Nov 2001 12:30 & 20 Nov 00:10 & 34\\
30$^{*}$ & 22 Nov 2001 23:20 & 24 Nov 05:55 & 18,900\\
31$^{*}$ & 26 Dec 2001 06:05 & 26 Dec 11:15 & 779\\
\hline
\multicolumn{4}{l}{$^{*}$ SGO stations affected by unknown radio interference.}
\end{tabular}
\label{tab:eventList1}
\end{table}

\begin{table}
\caption{\textit{Solar Proton Events Used in This Study (continued from previous table)}}
\centering
\begin{tabular}{lccc}
\hline
Event & SPE start time & SPE max time & Max $>10$ MeV flux \\ 
 & (UT) & (UT) & (pfu) \\ 
\hline
32$^{*}$ & 29 Dec 2001 05:10 & 29 Dec 08:15 & 76\\
33$^{*}$ & 30 Dec 2001 02:45 & 31 Dec 16:20 & 108\\
34$^{*}$ & 10 Jan 2002 20:45 & 11 Jan 05:30 & 91\\
35$^{*}$ & 15 Jan 2002 14:35 & 15 Jan 20:00 & 15\\
36$^{*}$ & 17 Mar 2002 08:20 & 17 Mar 08:50 & 13 \\
37$^{*}$ & 18 Mar 2002 13:00 & 19 Mar 06:50 & 53 \\
38$^{*}$ & 20 Mar 2002 15:10 & 20 Mar 15:25 & 19 \\
39$^{*}$ & 22 Mar 2002 20:20 & 23 Mar 13:20 & 16 \\
40 & 17 Apr 2002 15:30 & 17 Apr 15:40 & 24 \\
41 & 21 Apr 2002 02:25 & 21 Apr 23:20 & 2,520 \\
42 & 22 May 2002 17:55 & 23 May 10:55 & 820 \\
43 & 07 Jul 2002 18:30 & 07 Jul 19:55 & 22 \\
44 & 16 Jul 2002 17:50 & 17 Jul 16:00 & 234 \\
45 & 19 Jul 2002 10:50 & 19 Jul 15:15 & 13 \\
46 & 22 Jul 2002 06:55 & 23 Jul 10:25 & 28 \\
47 & 14 Aug 2002 09:00 & 14 Aug 16:20 & 26 \\
48 & 22 Aug 2002 04:40 & 22 Aug 09:40 & 36 \\
49 & 24 Aug 2002 01:40 & 24 Aug 08:35 & 317 \\
50 & 07 Sep 2002 04:40 & 07 Sep 16:50 & 208 \\
51 & 09 Nov 2002 19:20 & 10 Nov 05:40 & 404 \\
52 & 04 Nov 2003 22:25 & 05 Nov 06:00 & 353 \\
53$^{*}$ & 21 Nov 2003 23:55 & 22 Nov 02:30 & 13 \\
54$^{*}$ & 02 Dec 2003 15:05 & 02 Dec 17:30 & 86 \\
55 & 25 Jul 2004 18:55 & 26 Jul 22:50 & 2,086 \\
56 & 01 Nov 2004 06:55 & 01 Nov 08:05 & 63 \\
57 & 07 Nov 2004 19:10 & 08 Nov 01:15 & 495 \\
58 & 16 Jan 2005 02:10 & 17 Jan 17:50 & 5,040 \\
59 & 14 May 2005 05:25 & 15 May 02:40 & 3,140 \\
60 & 14 Jul 2005 02:45 & 15 Jul 03:45 & 134 \\
61 & 27 Jul 2005 23:00 & 29 Jul 17:15 & 41 \\
62 & 22 Aug 2005 20:40 & 23 Aug 10:45 & 330 \\
\hline
\multicolumn{4}{l}{$^{*}$ SGO stations affected by unknown radio interference.}
\end{tabular}
\label{tab:eventList2}
\end{table}

\acknowledgments
The work of E.H., P.T.V. and N.K. was supported by the Academy of Finland through the project \#276926 (SECTIC: Sun-Earth Connection Through Ion Chemistry). This work of A.K. is a part of the Tenure Track Project in Radio Science at Sodankyl\"a Geophysical Observatory. This work of N.P. was supported by the Research Council of Norway under CoE contract 223252. 
We acknowledge Peter Stauning from the Danish Meteorological Institute (Denmark) and Hisao Yamagishi from the National Institute of Polar Research (Japan) who provided the Longyearbyen imaging riometer data. 
The Kilpisj\"arvi riometer data originated from the Imaging Riometer for Ionospheric Studies (IRIS), operated by the Space Plasma Environment and Radio Science (SPEARS) group, Department of Physics, Lancaster University (UK) in collaboration with the Sodankyl\"a Geophysical Observatory. 
The Sodankyl\"a Geophysical Observatory riometer chain data were provided by Antti Kero and the Sodankyl\"a Geophysical Observatory (Finland). 
The GO-Canada riometer array is operated by the University of Calgary with financial support from the Canadian Space Agency. All GO-Canada riometer data are openly available from \texttt{data.phys.ucalgary.ca}. 
NOAA GOES particle flux data are openly available online from \texttt{https://www.ngdc.noaa.gov/stp/satellite/goes/dataaccess.html}. 
The authors would also like to acknowledge the openly available jLab data analysis package for Matlab~\citep{lilly2017}, which was used for parts of the plotting in this paper.


\begin{thebibliography}{50}
\providecommand{\natexlab}[1]{#1}
\expandafter\ifx\csname urlstyle\endcsname\relax
  \providecommand{\doi}[1]{doi:\discretionary{}{}{}#1}\else
  \providecommand{\doi}{doi:\discretionary{}{}{}\begingroup
  \urlstyle{rm}\Url}\fi

\bibitem[{\textit{Andersson et~al.}(2014)\textit{Andersson, Verronen, Rodger,
  Clilverd, and Sepp{\"a}l{\"a}}}]{andersson2014}
Andersson, M.~E., P.~T. Verronen, C.~J. Rodger, M.~A. Clilverd, and
  A.~Sepp{\"a}l{\"a} (2014), {Missing driver in the Sun--Earth connection from
  energetic electron precipitation impacts mesospheric ozone}, \textit{Nature
  Communications}, \textit{5}(5197), \doi{10.1038/ncomms6197}.

\bibitem[{\textit{Andersson et~al.}(2016)\textit{Andersson, Verronen, Marsh,
  P{\"a}iv{\"a}rinta, and Plane}}]{andersson2016}
Andersson, M.~E., P.~T. Verronen, D.~R. Marsh, S.-M. P{\"a}iv{\"a}rinta, and
  J.~M.~C. Plane (2016), {WACCM-D---Improved modeling of nitric acid and active
  chlorine during energetic particle precipitation}, \textit{Journal of
  Geophysical Research: Atmospheres}, \textit{121}(17), 10,328--10,341,
  \doi{10.1002/2015JD024173}.

\bibitem[{\textit{Banks and Kockarts}(1973)}]{banks1973}
Banks, P.~M., and G.~Kockarts (1973), \textit{Aeronomy}, vol.~A, Elsevier, New
  York, \doi{10.1016/C2013-0-10328-5}.

\bibitem[{\textit{Baumgaertner et~al.}(2011)\textit{Baumgaertner, Sepp\"al\"a,
  J\"ockel, and Clilverd}}]{baumgaertner2011}
Baumgaertner, A. J.~G., A.~Sepp\"al\"a, P.~J\"ockel, and M.~A. Clilverd (2011),
  {Geomagnetic activity related NO$_{x}$ enhancements and polar surface air
  temperature variability in a chemistry climate model: modulation of the NAM
  index}, \textit{Atmospheric Chemistry and Physics}, \textit{11}(9),
  4521--4531, \doi{10.5194/acp-11-4521-2011}.

\bibitem[{\textit{Browne et~al.}(1995)\textit{Browne, Hargreaves, and
  Honary}}]{browne1995}
Browne, S., J.~Hargreaves, and B.~Honary (1995), An imaging riometer for
  ionospheric studies, \textit{Electronics \& Communication Engineering
  Journal}, \textit{7}(5), 209--217, \doi{10.1049/ecej:19950505}.

\bibitem[{\textit{Clilverd et~al.}(2007)\textit{Clilverd, Rodger,
  Moffat-Griffin, and Verronen}}]{clilverd2007}
Clilverd, M.~A., C.~J. Rodger, T.~Moffat-Griffin, and P.~T. Verronen (2007),
  Improved dynamic geomagnetic rigidity cutoff modeling: Testing predictive
  accuracy, \textit{Journal of Geophysical Research: Space Physics},
  \textit{112}, A08302, \doi{10.1029/2007JA012410}.

\bibitem[{\textit{{Collis} and {Rietveld}}(1990)}]{collis1990}
{Collis}, P.~N., and M.~T. {Rietveld} (1990), {Mesospheric observations with
  the EISCAT UHF radar during polar cap absorption events: 1. Electron
  densities and negative ions}, \textit{Annales Geophysicae}, \textit{8},
  809--824.

\bibitem[{\textit{Friedrich et~al.}(2002)\textit{Friedrich, Harrich, Torkar,
  and Stauning}}]{friedrich2002}
Friedrich, M., M.~Harrich, K.~Torkar, and P.~Stauning (2002), Quantitative
  measurements with wide-beam riometers, \textit{Journal of Atmospheric and
  Solar-Terrestrial Physics}, \textit{64}(3), 359--365,
  \doi{10.1016/S1364-6826(01)00108-0}.

\bibitem[{\textit{Funke et~al.}(2014)\textit{Funke, L{\'o}pez-Puertas, Stiller,
  and von Clarmann}}]{funke2014}
Funke, B., M.~L{\'o}pez-Puertas, G.~P. Stiller, and T.~von Clarmann (2014),
  {Mesospheric and stratospheric NO$_y$ produced by energetic particle
  precipitation during 2002--2012}, \textit{Journal of Geophysical Research:
  Atmospheres}, \textit{119}(7), 4429--4446, \doi{10.1002/2013JD021404}.

\bibitem[{\textit{Gillett and Thompson}(2003)}]{gillett2003}
Gillett, N.~P., and D.~W.~J. Thompson (2003), {Simulation of recent Southern
  Hemisphere climate change}, \textit{Science}, \textit{302}(5643), 273--275,
  \doi{10.1126/science.1087440}.

\bibitem[{\textit{Hargreaves et~al.}(1987)\textit{Hargreaves, Ranta, Ranta,
  Turunen, and Turunen}}]{hargreaves1987}
Hargreaves, J., H.~Ranta, A.~Ranta, E.~Turunen, and T.~Turunen (1987),
  {Observations of the polar cap absorption event of February 1984 by the
  EISCAT incoherent scatter radar}, \textit{Planetary and Space Science},
  \textit{35}(7), 947--958, \doi{10.1016/0032-0633(87)90072-9}.

\bibitem[{\textit{Hargreaves and Birch}(2005)}]{hargreaves2005}
Hargreaves, J.~K., and M.~J. Birch (2005), {On the relations between proton
  influx and D-region electron densities during the polar-cap absorption event
  of 28--29 October 2003}, \textit{Annales Geophysicae}, \textit{23}(10),
  3267--3276, \doi{10.5194/angeo-23-3267-2005}.

\bibitem[{\textit{Hargreaves and Detrick}(2002)}]{hargreaves2002}
Hargreaves, J.~K., and D.~L. Detrick (2002), Application of polar cap
  absorption events to the calibration of riometer systems, \textit{Radio
  Science}, \textit{37}(3), 1--11, \doi{10.1029/2001RS002465}.

\bibitem[{\textit{Hedin}(1991)}]{hedin1991}
Hedin, A.~E. (1991), {Extension of the MSIS Thermosphere Model into the middle
  and lower atmosphere}, \textit{Journal of Geophysical Research: Space
  Physics}, \textit{96}(A2), 1159--1172, \doi{10.1029/90JA02125}.

\bibitem[{\textit{Jackman}(2013)}]{jackman2013}
Jackman, C.~H. (2013), Ionization rates for 1963--2012 from solar proton
  events, Retrieved from
  http://solarisheppa.geomar.de/solarisheppa/sites/default/files/data/SOLARIS\_Jackman\_SPEs.pdf.

\bibitem[{\textit{Jackman et~al.}(2005)\textit{Jackman, DeLand, Labow, Fleming,
  Weisenstein, Ko, Sinnhuber, Anderson, and Russell}}]{jackman2005}
Jackman, C.~H., M.~T. DeLand, G.~J. Labow, E.~L. Fleming, D.~K. Weisenstein,
  M.~K. Ko, M.~Sinnhuber, J.~Anderson, and J.~M. Russell (2005), The influence
  of the several very large solar proton events in years 2000--2003 on the
  neutral middle atmosphere, \textit{Advances in Space Research},
  \textit{35}(3), 445--450, \doi{10.1016/j.asr.2004.09.006}.

\bibitem[{\textit{Jackman et~al.}(2009)\textit{Jackman, Marsh, Vitt, Garcia,
  Randall, Fleming, and Frith}}]{jackman2009}
Jackman, C.~H., D.~R. Marsh, F.~M. Vitt, R.~R. Garcia, C.~E. Randall, E.~L.
  Fleming, and S.~M. Frith (2009), Long-term middle atmospheric influence of
  very large solar proton events, \textit{Journal of Geophysical Research:
  Atmospheres}, \textit{114}, D11304, \doi{10.1029/2008JD011415}.

\bibitem[{\textit{Jackman et~al.}(2016)\textit{Jackman, Marsh, Kinnison,
  Mertens, and Fleming}}]{jackman2016}
Jackman, C.~H., D.~R. Marsh, D.~E. Kinnison, C.~J. Mertens, and E.~L. Fleming
  (2016), Atmospheric changes caused by galactic cosmic rays over the period
  1960--2010, \textit{Atmospheric Chemistry and Physics}, \textit{16}(9),
  5853--5866, \doi{10.5194/acp-16-5853-2016}.

\bibitem[{\textit{{Kallenrode}}(2003)}]{kallenrode2003}
{Kallenrode}, M.-B. (2003), {Current views on impulsive and gradual solar
  energetic particle events}, \textit{Journal of Physics G: Nuclear Physics},
  \textit{29}, 965--981, \doi{10.1088/0954-3899/29/5/316}.

\bibitem[{\textit{{Kavanagh} et~al.}(2004)\textit{{Kavanagh}, {Marple},
  {Honary}, {McCrea}, and {Senior}}}]{kavanagh2004}
{Kavanagh}, A., S.~{Marple}, F.~{Honary}, I.~{McCrea}, and A.~{Senior} (2004),
  {On solar protons and polar cap absorption: Constraints on an empirical
  relationship}, \textit{Annales Geophysicae}, \textit{22}, 1133--1147,
  \doi{10.5194/angeo-22-1133-2004}.

\bibitem[{\textit{Kunz et~al.}(2011)\textit{Kunz, Pan, Konopka, Kinnison, and
  Tilmes}}]{kunz2011}
Kunz, A., L.~L. Pan, P.~Konopka, D.~E. Kinnison, and S.~Tilmes (2011),
  {Chemical and dynamical discontinuity at the extratropical tropopause based
  on START08 and WACCM analyses}, \textit{Journal of Geophysical Research:
  Atmospheres}, \textit{116}, D24302, \doi{10.1029/2011JD016686}.

\bibitem[{\textit{Lilly}(2017)}]{lilly2017}
Lilly, J.~M. (2017), {jLab: A data analysis package for Matlab, v. 1.6.5},
  Retrieved from http://www.jmlilly.net/jmlsoft.html.

\bibitem[{\textit{Little and Leinbach}(1958)}]{little1958}
Little, C.~G., and H.~Leinbach (1958), Some measurements of high-latitude
  ionospheric absorption using extraterrestrial radio waves,
  \textit{Proceedings of the IRE}, \textit{46}(1), 334--348,
  \doi{10.1109/JRPROC.1958.286795}.

\bibitem[{\textit{Little and Leinbach}(1959)}]{little1959}
Little, C.~G., and H.~Leinbach (1959), {The Riometer---A device for the
  continuous measurement of ionospheric absorption}, \textit{Proceedings of the
  IRE}, \textit{47}(2), 315--320, \doi{10.1109/JRPROC.1959.287299}.

\bibitem[{\textit{Marsh et~al.}(2007)\textit{Marsh, Garcia, Kinnison, Boville,
  Sassi, Solomon, and Matthes}}]{marsh2007}
Marsh, D.~R., R.~R. Garcia, D.~E. Kinnison, B.~A. Boville, F.~Sassi, S.~C.
  Solomon, and K.~Matthes (2007), Modeling the whole atmosphere response to
  solar cycle changes in radiative and geomagnetic forcing, \textit{Journal of
  Geophysical Research: Atmospheres}, \textit{112}, D23306,
  \doi{10.1029/2006JD008306}.

\bibitem[{\textit{Matthes et~al.}(2017)\textit{Matthes, Funke, Andersson,
  Barnard, Beer, Charbonneau, Clilverd, Dudok~de Wit, Haberreiter, Hendry,
  Jackman, Kretzschmar, Kruschke, Kunze, Langematz, Marsh, Maycock, Misios,
  Rodger, Scaife, Sepp\"al\"a, Shangguan, Sinnhuber, Tourpali, Usoskin, van~de
  Kamp, Verronen, and Versick}}]{matthes2017}
Matthes, K., B.~Funke, M.~E. Andersson, L.~Barnard, J.~Beer, P.~Charbonneau,
  M.~A. Clilverd, T.~Dudok~de Wit, M.~Haberreiter, A.~Hendry, C.~H. Jackman,
  M.~Kretzschmar, T.~Kruschke, M.~Kunze, U.~Langematz, D.~R. Marsh, A.~C.
  Maycock, S.~Misios, C.~J. Rodger, A.~A. Scaife, A.~Sepp\"al\"a, M.~Shangguan,
  M.~Sinnhuber, K.~Tourpali, I.~Usoskin, M.~van~de Kamp, P.~T. Verronen, and
  S.~Versick (2017), {Solar forcing for CMIP6 (v3.2)}, \textit{Geoscientific
  Model Development}, \textit{10}(6), 2247--2302,
  \doi{10.5194/gmd-10-2247-2017}.

\bibitem[{\textit{Nesse~Tyss{\o}y and Stadsnes}(2015)}]{nesse-tyssoy2015}
Nesse~Tyss{\o}y, H., and J.~Stadsnes (2015), {Cutoff latitude variation during
  solar proton events: Causes and consequences}, \textit{Journal of Geophysical
  Research: Space Physics}, \textit{120}(1), 553--563,
  \doi{10.1002/2014JA020508}.

\bibitem[{\textit{Nesse~Tyss{\o}y et~al.}(2013)\textit{Nesse~Tyss{\o}y,
  Stadsnes, S{\o}raas, and S{\o}rb{\o}}}]{nesse-tyssoy2013}
Nesse~Tyss{\o}y, H., J.~Stadsnes, F.~S{\o}raas, and M.~S{\o}rb{\o} (2013),
  {Variations in cutoff latitude during the January 2012 solar proton event and
  implication for the distribution of particle energy deposition},
  \textit{Geophysical Research Letters}, \textit{40}(16), 4149--4153,
  \doi{10.1002/grl.50815}.

\bibitem[{\textit{Randall et~al.}(2005)\textit{Randall, Harvey, Manney,
  Orsolini, Codrescu, Sioris, Brohede, Haley, Gordley, Zawodny, and
  Russell}}]{randall2005}
Randall, C.~E., V.~L. Harvey, G.~L. Manney, Y.~Orsolini, M.~Codrescu,
  C.~Sioris, S.~Brohede, C.~S. Haley, L.~L. Gordley, J.~M. Zawodny, and J.~M.
  Russell (2005), Stratospheric effects of energetic particle precipitation in
  2003--2004, \textit{Geophysical Research Letters}, \textit{32}, L05802,
  \doi{10.1029/2004GL022003}.

\bibitem[{\textit{Reames}(1999)}]{reames1999}
Reames, D.~V. (1999), {Particle acceleration at the Sun and in the
  heliosphere}, \textit{Space Science Reviews}, \textit{90}(3), 413--491,
  \doi{10.1023/A:1005105831781}.

\bibitem[{\textit{Rienecker et~al.}(2011)\textit{Rienecker, Suarez, Gelaro,
  Todling, Bacmeister, Liu, Bosilovich, Schubert, Takacs, Kim, Bloom, Chen,
  Collins, Conaty, da~Silva, Gu, Joiner, Koster, Lucchesi, Molod, Owens,
  Pawson, Pegion, Redder, Reichle, Robertson, Ruddick, Sienkiewicz, and
  Woollen}}]{rienecker2011}
Rienecker, M.~M., M.~J. Suarez, R.~Gelaro, R.~Todling, J.~Bacmeister, E.~Liu,
  M.~G. Bosilovich, S.~D. Schubert, L.~Takacs, G.-K. Kim, S.~Bloom, J.~Chen,
  D.~Collins, A.~Conaty, A.~da~Silva, W.~Gu, J.~Joiner, R.~D. Koster,
  R.~Lucchesi, A.~Molod, T.~Owens, S.~Pawson, P.~Pegion, C.~R. Redder,
  R.~Reichle, F.~R. Robertson, A.~G. Ruddick, M.~Sienkiewicz, and J.~Woollen
  (2011), {MERRA: NASA's Modern-Era Retrospective Analysis for Research and
  Applications}, \textit{Journal of Climate}, \textit{24}(14), 3624--3648,
  \doi{10.1175/JCLI-D-11-00015.1}.

\bibitem[{\textit{Rodger et~al.}(2006)\textit{Rodger, Clilverd, Verronen,
  Ulich, Jarvis, and Turunen}}]{rodger2006}
Rodger, C.~J., M.~A. Clilverd, P.~T. Verronen, T.~Ulich, M.~J. Jarvis, and
  E.~Turunen (2006), Dynamic geomagnetic rigidity cutoff variations during a
  solar proton event, \textit{Journal of Geophysical Research: Space Physics},
  \textit{111}, A04222, \doi{10.1029/2005JA011395}.

\bibitem[{\textit{Rodger et~al.}(2010)\textit{Rodger, Clilverd,
  Sepp{\"a}l{\"a}, Thomson, Gamble, Parrot, Sauvaud, and Ulich}}]{rodger2010}
Rodger, C.~J., M.~A. Clilverd, A.~Sepp{\"a}l{\"a}, N.~R. Thomson, R.~J. Gamble,
  M.~Parrot, J.-A. Sauvaud, and T.~Ulich (2010), Radiation belt electron
  precipitation due to geomagnetic storms: Significance to middle atmosphere
  ozone chemistry, \textit{Journal of Geophysical Research: Space Physics},
  \textit{115}, A11320, \doi{10.1029/2010JA015599}.

\bibitem[{\textit{Rogers et~al.}(2016)\textit{Rogers, Kero, Honary, Verronen,
  Warrington, and Danskin}}]{rogers2016}
Rogers, N.~C., A.~Kero, F.~Honary, P.~T. Verronen, E.~M. Warrington, and D.~W.
  Danskin (2016), Improving the twilight model for polar cap absorption
  nowcasts, \textit{Space Weather}, \textit{14}(11), 950--972,
  \doi{10.1002/2016SW001527}.

\bibitem[{\textit{Rosenberg et~al.}(1991)\textit{Rosenberg, Detrick,
  Venkatesan, and van Bavel}}]{rosenberg1991}
Rosenberg, T.~J., D.~L. Detrick, D.~Venkatesan, and G.~van Bavel (1991), A
  comparative study of imaging and broad-beam riometer measurements: The effect
  of spatial structure on the frequency dependence of auroral absorption,
  \textit{Journal of Geophysical Research: Space Physics}, \textit{96}(A10),
  17,793--17,803, \doi{10.1029/91JA01827}.

\bibitem[{\textit{Rostoker et~al.}(1995)\textit{Rostoker, Samson, Creutzberg,
  Hughes, McDiarmid, McNamara, Jones, Wallis, and Cogger}}]{rostoker1995}
Rostoker, G., J.~C. Samson, F.~Creutzberg, T.~J. Hughes, D.~R. McDiarmid, A.~G.
  McNamara, A.~V. Jones, D.~D. Wallis, and L.~L. Cogger (1995), {Canopus --- A
  ground-based instrument array for remote sensing the high latitude ionosphere
  during the ISTP/GGS program}, \textit{Space Science Reviews}, \textit{71}(1),
  743--760, \doi{10.1007/BF00751349}.

\bibitem[{\textit{Sen and Wyller}(1960)}]{sen1960}
Sen, H.~K., and A.~A. Wyller (1960), {On the generalization of the
  Appleton-Hartree magnetoionic formulas}, \textit{Journal of Geophysical
  Research}, \textit{65}(12), 3931--3950, \doi{10.1029/JZ065i012p03931}.

\bibitem[{\textit{Sepp{\"a}l{\"a} et~al.}(2004)\textit{Sepp{\"a}l{\"a},
  Verronen, Kyr{\"o}l{\"a}, Hassinen, Backman, Hauchecorne, Bertaux, and
  Fussen}}]{seppala2004}
Sepp{\"a}l{\"a}, A., P.~T. Verronen, E.~Kyr{\"o}l{\"a}, S.~Hassinen,
  L.~Backman, A.~Hauchecorne, J.~L. Bertaux, and D.~Fussen (2004), {Solar
  proton events of October--November 2003: Ozone depletion in the Northern
  Hemisphere polar winter as seen by GOMOS/Envisat}, \textit{Geophysical
  Research Letters}, \textit{31}, L19107, \doi{10.1029/2004GL021042}.

\bibitem[{\textit{Sepp{\"a}l{\"a} et~al.}(2009)\textit{Sepp{\"a}l{\"a},
  Randall, Clilverd, Rozanov, and Rodger}}]{seppala2009}
Sepp{\"a}l{\"a}, A., C.~E. Randall, M.~A. Clilverd, E.~Rozanov, and C.~J.
  Rodger (2009), Geomagnetic activity and polar surface air temperature
  variability, \textit{Journal of Geophysical Research: Space Physics},
  \textit{114}, A10312, \doi{10.1029/2008JA014029}.

\bibitem[{\textit{Sepp{\"a}l{\"a} et~al.}(2015)\textit{Sepp{\"a}l{\"a},
  Clilverd, Beharrell, Rodger, Verronen, Andersson, and Newnham}}]{seppala2015}
Sepp{\"a}l{\"a}, A., M.~A. Clilverd, M.~J. Beharrell, C.~J. Rodger, P.~T.
  Verronen, M.~E. Andersson, and D.~A. Newnham (2015), Substorm-induced
  energetic electron precipitation: Impact on atmospheric chemistry,
  \textit{Geophysical Research Letters}, \textit{42}(19), 8172--8176,
  \doi{10.1002/2015GL065523}.

\bibitem[{\textit{Sinnhuber et~al.}(2012)\textit{Sinnhuber, Nieder, and
  Wieters}}]{sinnhuber2012}
Sinnhuber, M., H.~Nieder, and N.~Wieters (2012), Energetic particle
  precipitation and the chemistry of the mesosphere/lower thermosphere,
  \textit{Surveys in Geophysics}, \textit{33}(6), 1281--1334,
  \doi{10.1007/s10712-012-9201-3}.

\bibitem[{\textit{Smith-Johnsen et~al.}(2018)\textit{Smith-Johnsen, Marsh,
  Orsolini, Nesse~Tyss{\o}y, Hendrickx, Sandanger, {\O}degaard, and
  Stordal}}]{smithjohnsen2018}
Smith-Johnsen, C., D.~R. Marsh, Y.~Orsolini, H.~Nesse~Tyss{\o}y, K.~Hendrickx,
  M.~I. Sandanger, L.-K.~G. {\O}degaard, and F.~Stordal (2018), {Nitric oxide
  response to the April 2010 electron precipitation event: Using WACCM and
  WACCM-D with and without medium energy electrons}, \textit{Journal of
  Geophysical Research: Space Physics}, \textit{123}(6), 5232--5245,
  \doi{10.1029/2018JA025418}.

\bibitem[{\textit{Spanswick et~al.}(2005)\textit{Spanswick, Donovan, and
  Baker}}]{spanswick2005}
Spanswick, E., E.~Donovan, and G.~Baker (2005), {Pc5 modulation of high energy
  electron precipitation: Particle interaction regions and scattering
  efficiency}, \textit{Annales Geophysicae}, \textit{23}(5), 1533--1542,
  \doi{10.5194/angeo-23-1533-2005}.

\bibitem[{\textit{Stauning and Hisao}(1995)}]{stauning1995}
Stauning, P., and Y.~Hisao (1995), {Imaging Riometer Installation in
  Longyearbyen, Svalbard}, \textit{Technical report 95-12}, Danish
  Meteorological Institute, Copenhagen.

\bibitem[{\textit{Vainio et~al.}(2009)\textit{Vainio, Desorgher, Heynderickx,
  Storini, Fl{\"u}ckiger, Horne, Kovaltsov, Kudela, Laurenza, McKenna-Lawlor,
  Rothkaehl, and Usoskin}}]{vainio2009}
Vainio, R., L.~Desorgher, D.~Heynderickx, M.~Storini, E.~Fl{\"u}ckiger, R.~B.
  Horne, G.~A. Kovaltsov, K.~Kudela, M.~Laurenza, S.~McKenna-Lawlor,
  H.~Rothkaehl, and I.~G. Usoskin (2009), {Dynamics of the Earth's particle
  radiation environment}, \textit{Space Science Reviews}, \textit{147}(3),
  187--231, \doi{10.1007/s11214-009-9496-7}.

\bibitem[{\textit{van~de Kamp et~al.}(2016)\textit{van~de Kamp,
  Sepp{\"a}l{\"a}, Clilverd, Rodger, Verronen, and Whittaker}}]{kamp2016}
van~de Kamp, M., A.~Sepp{\"a}l{\"a}, M.~A. Clilverd, C.~J. Rodger, P.~T.
  Verronen, and I.~C. Whittaker (2016), A model providing long-term data sets
  of energetic electron precipitation during geomagnetic storms,
  \textit{Journal of Geophysical Research: Atmospheres}, \textit{121}(20),
  12,520--12,540, \doi{10.1002/2015JD024212}.

\bibitem[{\textit{van~de Kamp et~al.}(2018)\textit{van~de Kamp, Rodger,
  Sepp{\"a}l{\"a}, Clilverd, and Verronen}}]{kamp2018}
van~de Kamp, M., C.~J. Rodger, A.~Sepp{\"a}l{\"a}, M.~A. Clilverd, and P.~T.
  Verronen (2018), An updated model providing long-term data sets of energetic
  electron precipitation, including zonal dependence, \textit{Journal of
  Geophysical Research: Atmospheres}, \textit{123}(17), 9891--9915,
  \doi{10.1029/2017JD028253}.

\bibitem[{\textit{Verronen and Lehmann}(2013)}]{verronen2013}
Verronen, P.~T., and R.~Lehmann (2013), {Analysis and parameterisation of ionic
  reactions affecting middle atmospheric HO$_x$ and NO$_y$ during solar proton
  events}, \textit{Annales Geophysicae}, \textit{31}(5), 909--956,
  \doi{10.5194/angeo-31-909-2013}.

\bibitem[{\textit{Verronen et~al.}(2006)\textit{Verronen, Ulich, Turunen, and
  Rodger}}]{verronen2006}
Verronen, P.~T., T.~Ulich, E.~Turunen, and C.~J. Rodger (2006), {Sunset
  transition of negative charge in the D-region ionosphere during
  high-ionization conditions}, \textit{Annales Geophysicae}, \textit{24}(1),
  187--202, \doi{10.5194/angeo-24-187-2006}.

\bibitem[{\textit{Verronen et~al.}(2016)\textit{Verronen, Andersson, Marsh,
  Kov{\'a}cs, and Plane}}]{verronen2016}
Verronen, P.~T., M.~E. Andersson, D.~R. Marsh, T.~Kov{\'a}cs, and J.~M.~C.
  Plane (2016), {WACCM-D---Whole Atmosphere Community Climate Model with
  D-region ion chemistry}, \textit{Journal of Advances in Modeling Earth
  Systems}, \textit{8}(2), 954--975, \doi{10.1002/2015MS000592}.

\end{thebibliography}
\end{document}